# A calibrated digital sideband separating spectrometer for radio astronomy applications

Ricardo Finger[1,2], Patricio Mena[2], Nicolás Reyes[1,2], Rafael Rodriguez[2], Leonardo Bronfman[1]

(1) Astronomy Department, Universidad de Chile, Camino El Observatorio 1515, Santiago, Chile
(2) Electrical Engineering Department, Universidad de Chile, Av. Tupper 2007, Santiago, Chile
Contact: rfinger@das.uchile.cl

## Abstract

Dual sideband (2SB) receivers are well suited for the spectral observation of complex astronomical signals over a wide frequency range. They are extensively used in radio astronomy, their main advantages being to avoid spectral confusion and to diminish effective system temperature by a factor two with respect to double sideband (DSB) receivers. Using available millimeter-wave analog technology, wideband 2SB receivers generally obtain sideband rejections ratios (SRR) of 10 to 15dB, insufficient for a number of astronomical applications. We report here the design and implementation of an FPGA-based sideband separating FFT spectrometer. A 4GHz analog front end was built to test the design and measure sideband rejection. The setup uses a 2SB front end architecture, except that the mixer outputs are directly digitized before the IF hybrid, using two 8bits ADCs sampling at 1GSPS. The IF hybrid is implemented on the FPGA together with a set of calibration vectors that, properly chosen, compensate for the analog front end amplitude and phase imbalances. The calibrated receiver exhibits a sideband rejection ratio in excess of 40dB for the entire 2GHz RF bandwidth.

## 1 Introduction

Much effort has been invested in the last decades improving the sensitivity of radio telescopes. Cryogenic heterodyne receivers are quickly approaching fundamental limits with respect to their noise temperature, but are still showing a relatively moderate performance in other parameters like Sideband Rejection Ratio (SRR). The Atacama Large Millimeter/Submillimeter Array (ALMA) specification for its state-of-the-art receivers requires an SRR better than 7dB over the entire band and better than 10dB over the 90% of the band (C. T. Cunningham *et al*, 2007). What seems to be a moderate specification proved to be difficult to achieve, particularly at the higher frequencies (F. P. Mena *et al*, 2011)( S. Mahieu, et al, 2011).

The preferred configuration of radio-astronomy heterodyne receivers is the dual sideband separating architecture (2SB), on which the two sidebands are output on different ports allowing simultaneous observations of cleaner and wider spectra. Sideband separating receivers are nevertheless difficult to build since they require the construction of two hybrids plus two parallel mixer/amplifier chains with excellent amplitude and phase balance over wide bandwidths.

Figure 1 shows the standard configuration of a 2SB front end. The first stage consists of an RF hybrid which splits the incoming -RF- signal in two paths with ideally 90° phase difference. The RF hybrid outputs are down-converted by the mixers $M_1$ and $M_2$, which are driven by the same Local Oscillator (LO) and recombined by the IF hybrid producing the outputs $I_1$, $I_2$. Alternatively, the RF input may be divided in-phase while the LO is divided in quadrature, as in Figure 2. For an ideal design the outputs $I_1$ and $I_2$ are the exact LSB and USB components of the incoming RF signal (Bert C. Henderson & James A. Cook, 1985). In real implementations the sideband separation is imperfect. Always part of the USB/LSB leaks into the unwanted port, with each output thus having a combination of both sidebands. The unavoidable gain and phase imbalances of the two parallel paths strongly limits the achievable sideband rejection ratio to about 10-20 dB for high-sensitivity, broadband receivers.

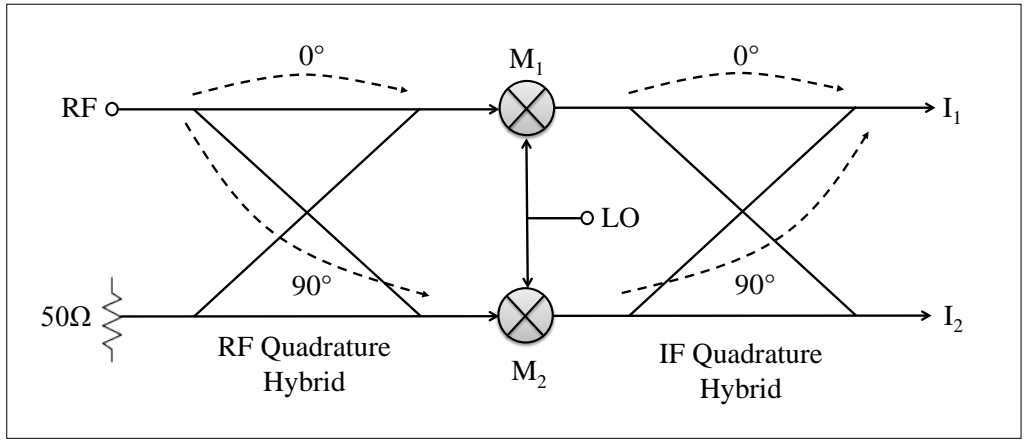

**Figure 1: Typical Sideband Separating Mixer configuration. Ideally the outputs $I_1$ and $I_2$ have only the LSB and USB components of the incoming RF signal respectively. Actual radio astronomy broadband receivers exhibit a sideband rejection ratio in the range of 10–20 dB.**

The increasing speed of digital hardware has opened the door for a new approach which is based on the idea of performing the IF recombination digitally. In this method, the mixer outputs are directly digitized using two analog to digital converters (ADC) so that a digital IF hybrid may be implemented which is not only ideal, but can also correct the imbalances of the analog RF and IF components with digital processing.

A receiver with digital sideband separation was reported in 2009 by Alex Murk *et al.* capable of processing 2×500MHz of bandwidth in real-time, using a Field Programmable Gate Array (FPGA). They reported calculated image rejections between 10 and 25 dB, which are at the high end of current analog sideband separating technology. Even though they did not use the digital back end to correct for imbalances of the RF front end, they clearly showed the applicability of FPGA based platforms to implement signal processes normally performed by analog hardware.

A prototype proving calibrated sideband separation has been recently (2010) constructed by M.A. Morgan and J.R. Fisher at NRAO. Figure 2 shows the architecture of Morgan and Fisher's Digital Sideband Separating Mixer (DSSM).

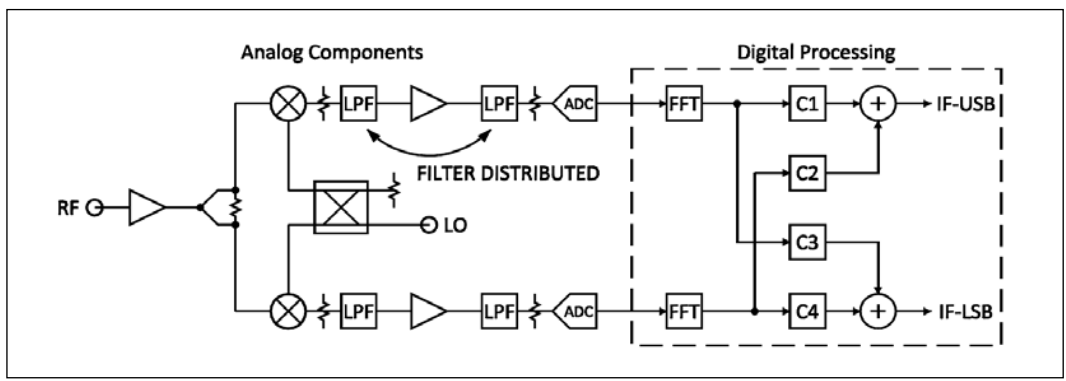

**Figure 2: Morgan and Fisher Digital Sideband Separating Mixer (DSSM). C1 to C4 are 1-dimesional complex arrays used to implement the IF hybrid and correct the amplitude and phase imbalances of the analog components. In this experiment the digital processing was performed *off-line*, using a desktop computer**

Their method consists of digitizing the mixer outputs and then calculating the Fourier transform. After the spectrum is calculated, each frequency channel (from both mixers) is duplicated and multiplied by the complex constants C1 to C4, which represent 1-dimesional complex arrays, of a length equal to the number of channels of the FFT. When C1 to C4 are chosen carefully one can not only reproduce digitally the behavior of an ideal IF hybrid but also calibrate out the accumulated RF and IF imbalances of both signal branches. In their experiment, Morgan and Fisher processed 2x250 MHz of bandwidth down-converted from L-Band (1.2-1.7GHz) with a sideband rejection ratio better than 50dB over the entire band, which represents an outstanding sideband separation 20 to 30dB better than current analog technology (M.A. Morgan & J.R. Fisher, 2010). Even though the digital processing shown in that experiment was performed *off-line,* using a desktop computer, it does show the feasibility of increasing the sideband rejection of current sideband separating receivers using digital technology.

In this report we present a next step from the digital sideband separation reported by A.Murk *et al*. and proof of concept of M.Morgan and J.Fisher by building a real-time, calibrated, sideband separating receiver using FPGA technology.

## 2   Experiment

### 2.1   Hardware description

Figure 3 shows the sideband separating spectrometer block diagram. A 4 GHz analog front end was built out of commercial parts to provide the functionality of typical analog receivers. Following the 90° RF hybrid two mixers down-convert the input signal using a 2.2–3.2-GHz LO. After amplification and anti-aliasing filtering the outputs are digitized to 8 bits, at 1 GSPS, allowing the processing of a 500 MHz IF bandwidth per sideband. Even though the RF frequency of this prototype is relatively low, there is nothing preventing the design to operate at higher RF frequencies provided a suitable front end.

The hardware used to perform the signal processing is known as the Reconfigurable Open Architecture Computing Hardware (ROACH) (CASPER wiki site, 2013). The ROACH is an open FPGA-based (Xilinx Virtex 5) platform, the product of an international collaboration lead by the Center for Astronomy Signal Processing and Electronic Research (CASPER) at the University of California, Berkeley. CASPER aims to produce open hardware designs and software/gateware resources for signal processing in astronomy (CASPER main web site, 2013).

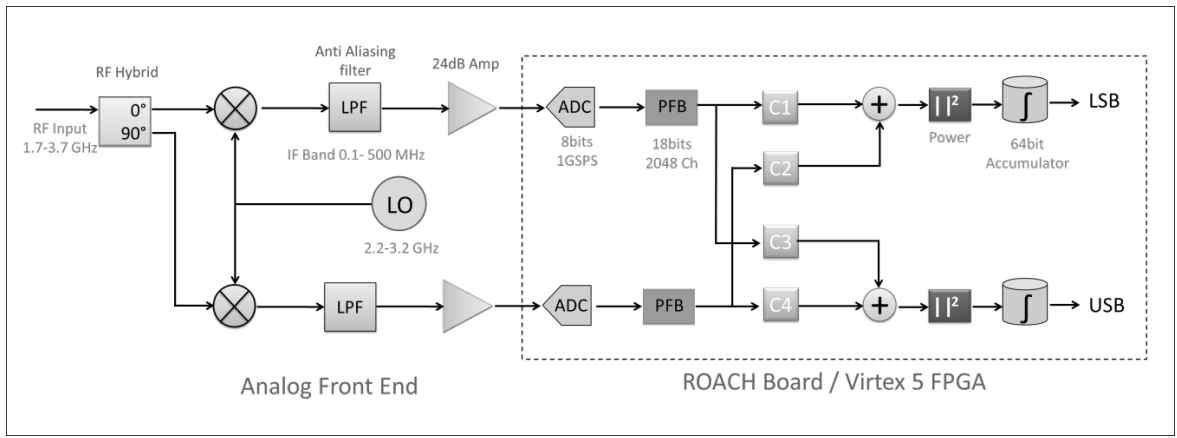

**Figure 3: Digital Sideband Separating Receiver Block Diagram. The digital processing and accumulation is performed in a Xilinx Virtex 5 FPGA.**

Our design closely resembles Morgan and Fisher's prototype but includes a power block and a 64 bit accumulator to integrate the high data rate output within the FPGA. Also, a simpler filtering scheme and in-phase LO injection were used in our front end. The complex constants C1 and C4 were set to 1, while C2 and C3 were adjusted to calibrate the phase and amplitude imbalances. C2 and C3 and the accumulation length parameter are accessible to the user and can be modified from the control computer anytime.

The core element of the spectrometer is the Polyphase Filter Bank (PFB) block, which is made of a finite impulse response filter followed by a pipeline Fast Fourier Transform. The pipeline FFT is an algorithm which can compute the FFT in a sequential manner. It achieves real-time behavior with nonstop processing when data is continually fed through the processor (Bin Zhou et al., 2007). The PFB has 2048 channels with data and coefficient bit-widths of 18bits. To build the PFB, the standard CASPER library IP blocks were used.

In a digital spectrometer the bandwidth is normally limited by the maximum ADC or by the logic (FPGA) clock speed. In our design, the limitation came first from the FPGA maximum clock speed that ensures the correct propagation of the signals within the FPGA fabric.

### 2.2 Test setup

Figure 4 shows a picture of the analog plate. Figure 5 shows the measuring test setup block diagram. Two signal synthesizers were used as LO and RF sources. A noise source was used to provide a white noise floor, which was coupled to the test tone to improve the ADC performance as described by M.A. Morgan & J.R. Fisher, 2010, and to emulate a typical radio astronomical application. The analog plate has two heaters to increase its temperature so the calibration thermal stability can be measured. No particular care was taken to match cable length or to choose matched-pairs of mixers, amplifiers or any other component. Both LO and RF synthesizers, as well as the clock generator for the ADCs and ROACH are locked to the same 10MHz reference.

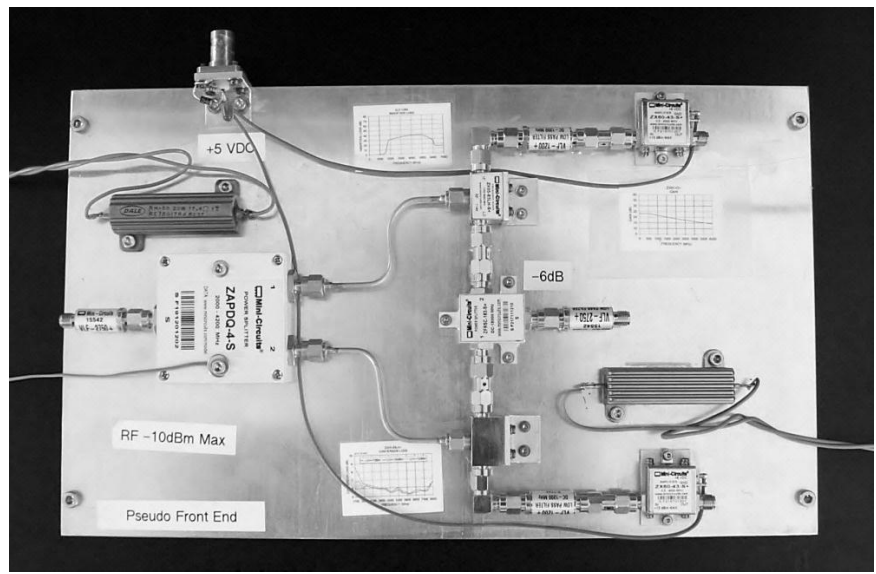

**Figure 4: Front End analog plate. From left to right it can be seen the RF input filter, RF quadrature hybrid, two semi-rigid cables leading to the mixers, anti-aliasing filters and amplifiers. In the center of the picture the LO input filter and LO splitter are seen. The filters are used to attenuate harmonics coming from the synthesizers. The darker corrugated components on both sides are the heaters.**

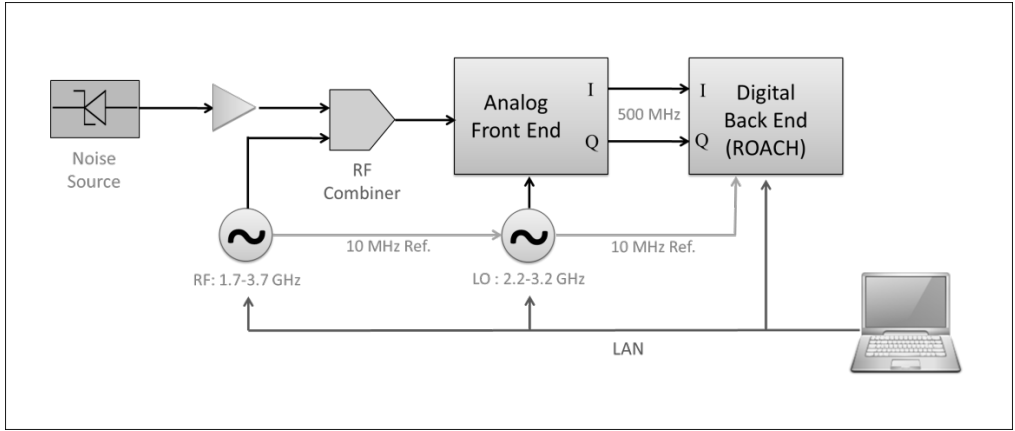

**Figure 5:** Test setup block diagram. Both LO and RF synthesizers, as well as the clock generator for the ADCs and ROACH are locked to the same 10MHz reference. A LAN is used to control the instrument and download the spectra.

## 3  Calibration

A second spectrometer was designed to run on the same setup which, instead of accumulating the sideband-separated power spectrum, records the amplitude and phase of 1024 spectral channels (even-numbered channels). This configuration directly stores the PFB complex outputs for further analysis on a desktop computer. A test tone is used to measure the amplitude and phase of both analog branches for the 1024 channels. The measurement takes 40 minutes to be completed. With the collected data the calibration constants C1-C4 are calculated in the following way: C1 and C4 are set to 1+0j while C2 and C3 are determined using the formulas:

and,
$$\frac{C_1}{C_2} = \frac{1}{X} e^{-j(\emptyset_{LSB} - \pi)} \tag{1}$$

$$\frac{C_3}{C_4} = \frac{1}{X} e^{-j(\emptyset_{USB} - \pi)} \tag{2}$$

where $X$ is the amplitude ratio of the two IF branches and $\emptyset_{LSB/USB}$ are the differential phase at the ADC input measured on each sideband (M.A. Morgan & J.R. Fisher, 2010). The odd-numbered channels calibration coefficients are calculated by linear interpolation of the even-numbered channels (measured data). Finally C1 to C4 are written into the FPGA, a process that takes less than a minute.

## 4  Results

### 4.1  Sideband Rejection Measurements

After calibration, the sideband rejection ratio is measured for every spectral channel by direct calculation of the amplitude ratio of the test tone in both sidebands. This approach can be used since the amplitude imbalance of both signal branches has been calibrated, so no additional correction to the direct SRR measurement is needed.

Figure 6 shows the USB and LSB spectra when a full scale (0dBm at the ADC input) test tone is applied on the USB (RF=2.6GHz, LO=2.5GHz). For this example an ideal (uncalibrated) digital hybrid was implemented (C1=C4=1+0j, C2=C3 =0+1j). Spurious signals generated mostly by the ADC are seen around -50dBc limiting the dynamic range of the setup to about 50 dB. The sideband rejection in the example exceeds 20dB.

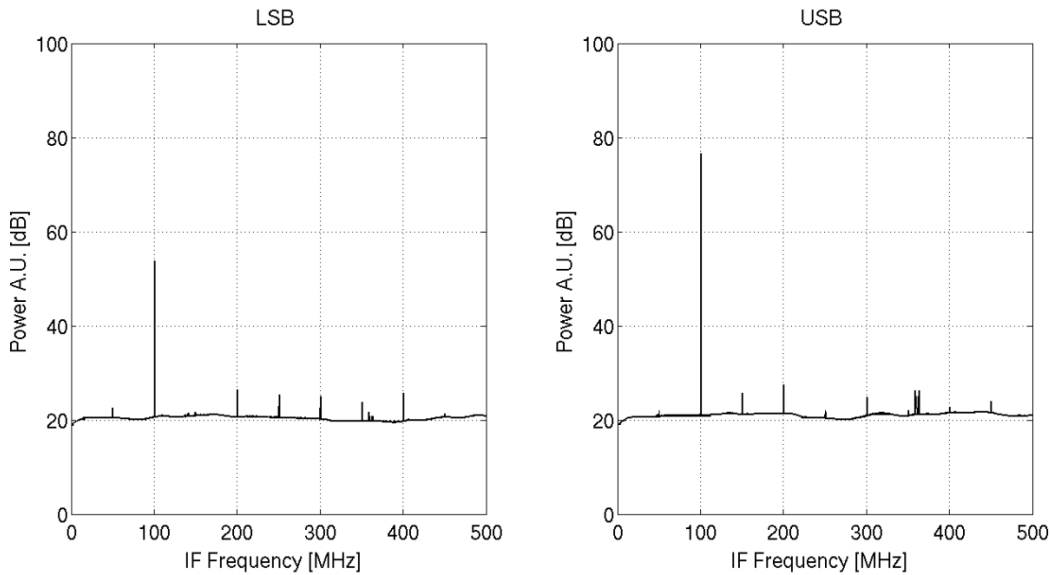

**Figure 6: Uncalibrated LSB and USB spectra for a 2.6GHz test tone. LO was set to 2.5GHz. Spurious signals generated in the ADC can be seen at -50dBc on the pass band (USB). The sideband rejection in the example exceeds 20dB.**

Figure 7 shows the amplitude and phase imbalances of the LSB and USB (LO is set at 2.5GHz). The data shown were taken at 31±1 °C and are used to calculate the constants C2 and C3.

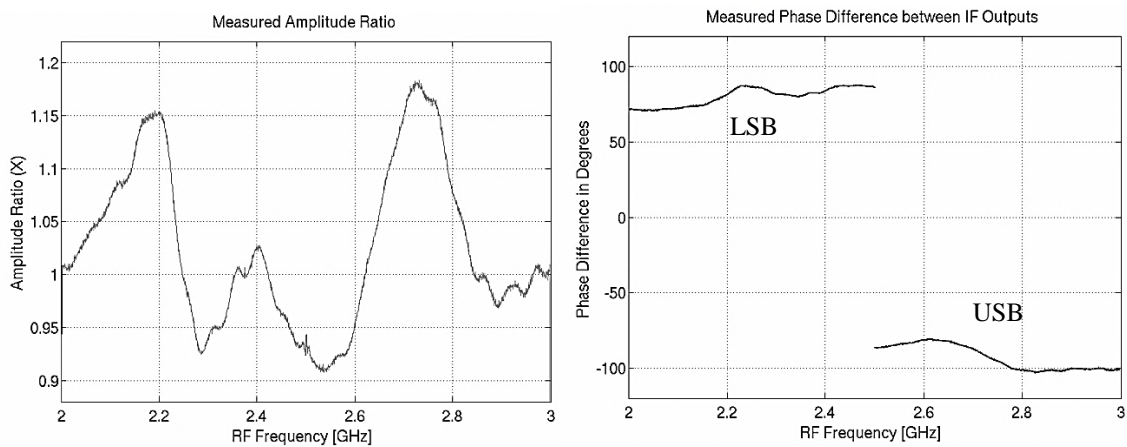

**Figure 7: Measured USB and LSB amplitude ratio *X* (left) and phase imbalances ∅<sub>LSB/USB</sub> (right). RF input from 2 to 3GHz, LO=2.5GHz**

Based on the data shown Figure 7 right, it is possible to calculate the phase unbalance contribution of the LO/RF and IF part independently. The calculation is performed using the expressions (J.R. Fisher & M.A. Morgan, 2008):

$$\emptyset_{LSB} = \emptyset_{IF} + \emptyset_{LO} \tag{3}$$

$$\emptyset_{USB} = \emptyset_{IF} - \emptyset_{LO} \tag{4}$$

where $\emptyset_{LO}$ represent the phase unbalance at the ADC input due to the RF and LO components, and $\emptyset_{IF}$ the signal path mismatch after the mixers. The values are shown in Figure 8.

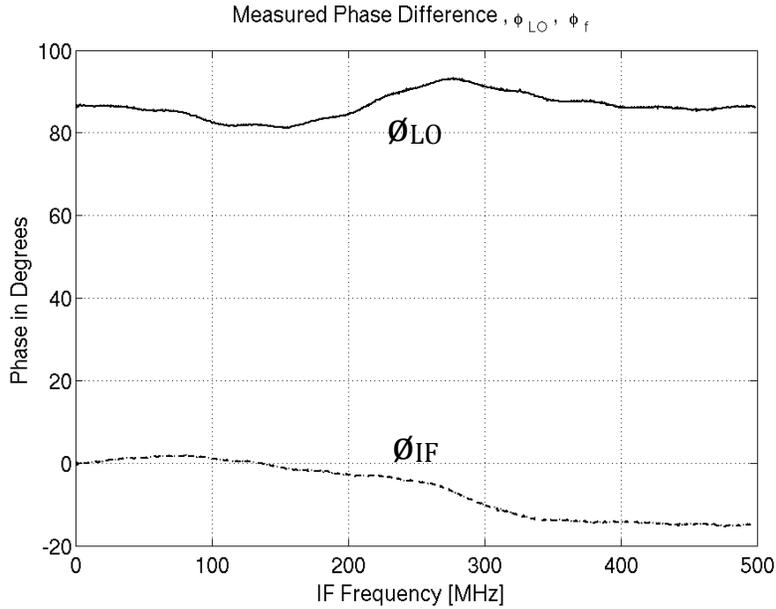

**Figure 8: Measured $\emptyset_{LO}$ (solid line) and $\emptyset_{IF}$ (dashed line) contribution to the total phase difference.**

Figure 9 shows the sideband rejection ratio for an RF input from 2 to 3GHz, LO=2.5GHz. The sideband rejection is better than 40dB for the entire band and better than 50dB for most of the band. The drop in SRR close to RF=2.5GHz is due to the DC decoupling of the ADCs. The calibration and measurements were performed with the analog front end at 31±1 °C. When the test tone amplitude on the rejected sideband is less than the spurious signal content, the SRR is measured as the ratio of the test tone in the pass band to the stronger spurious in the rejected band, so the measurement is limited by the ADC spurious-free dynamic range to about 50dB.

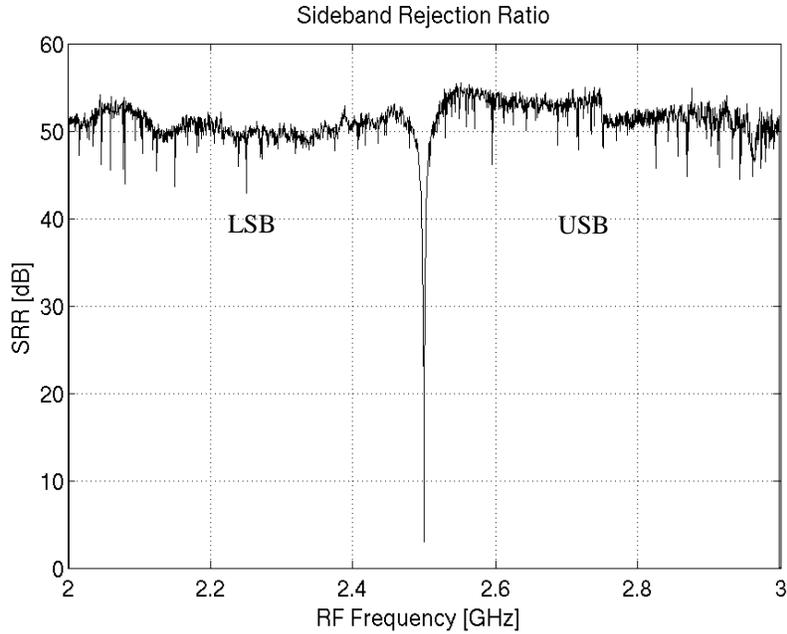

**Figure 9: Sideband Rejection Ratio after calibration of all spectral channels. The drop in SRR close to RF=2.5GHz is due to the DC decoupling of the ADCs. The lower values closed to 45 dB are associated with stronger spurious signals produced by the ADC.**

Figure 10 shows the sideband rejection ratio for an RF band of 1.7 to 3.7 GHz. Two LOs were used at 2.2 and 3.2 GHz. The figure shows that the sideband rejection is similar for different LOs and shows no degradation on the edges of the RF band.

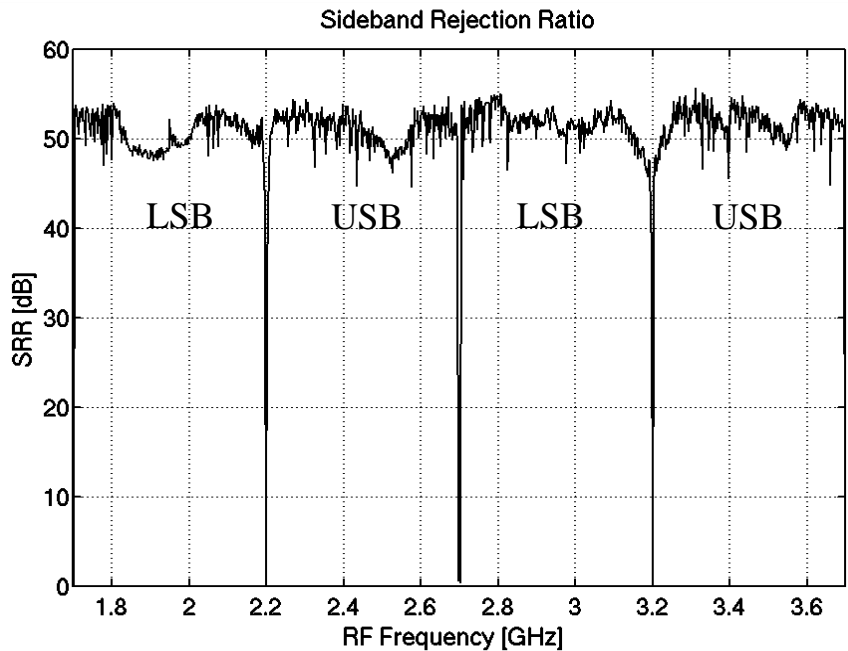

**Figure 10: Two LO´s Sideband Rejection Ratio. LO were set to 2.2 and 3.2 GHz while RF ranged from 1.7 to 3.7 GHz. 512 channels per sideband were measured for this figure.**

## 4.2 Thermal stability measurements

To study the thermal stability of the calibration the analog plate was heated up to 40±1 °C and the SRR was measured for 512 channels. The calibration coefficients remain the same as that measured at 31±1 °C. Figure 11 shows a degradation of up to 20dB for this experiment.

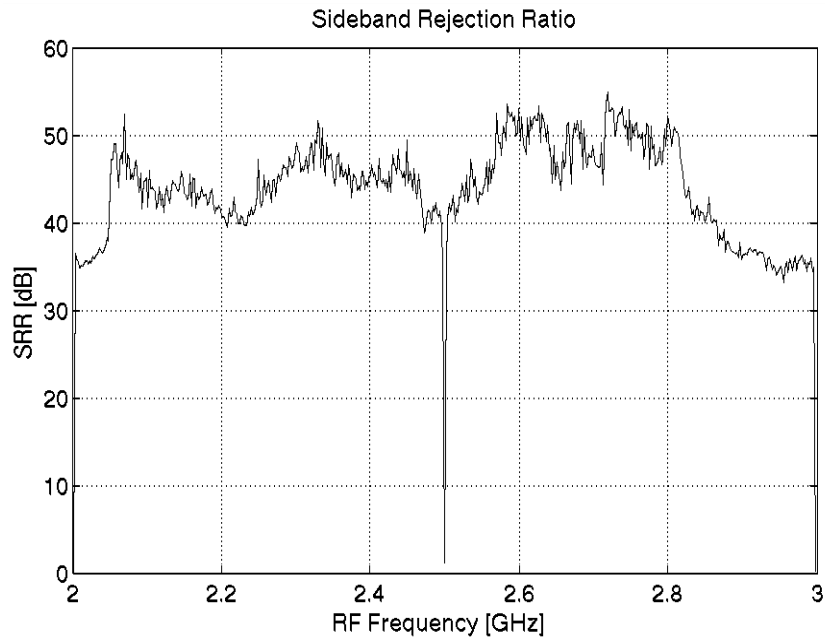

**Figure 11: Sideband Rejection Ratio measured at 40±1C with calibration performed at 31±1C. The figure shows the calibration thermal stability of the analog front end used for this experiment. Shorter cables and higher integration particularly of the RF and LO components may improve the thermal stability.**

The main cause of SRR sensitivity is the temperature induced phase unbalance due to the low integration of the test front end. Our front end is made of connectorized components mounted on rather big (31x18cm) plate. Also two 9cm cables were used between the RF hybrid and the mixers. Better integration, particularly of the RF and LO components, should improve SRR thermal stability.

After cooling back to 31±1°C the SRR returns to the essentially the same values of Figure 9 suggesting a good long term stability of the calibration. Three such thermal cycles were performed within 48 hours. The SRR was measured for each cycle at 31±1°C. Results are shown in Figure 12.

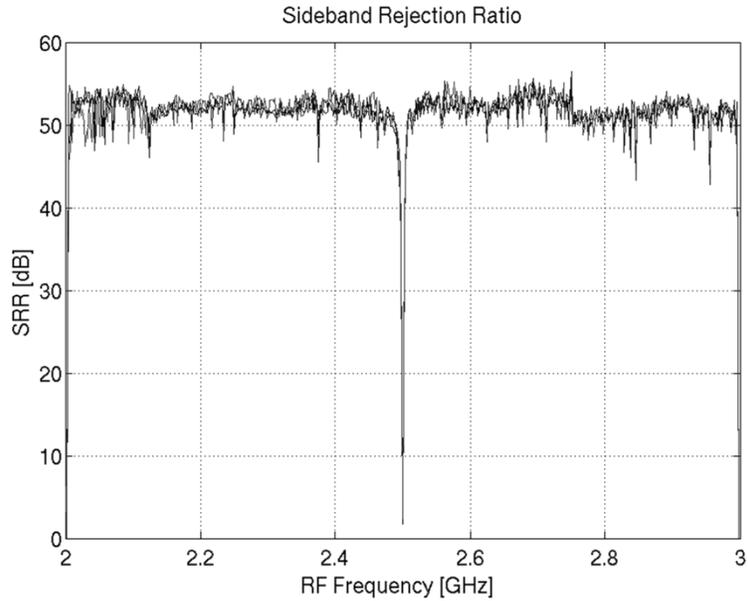

**Figure 12: Sideband Rejection Ratio measured at 31±1°C for 3 thermal cycles. The plate was heated up to 40±1°C every time. The calibration was performed only once prior the thermal cycling. The curves overlap almost perfectly showing a good stability of the calibration.**

## 5  Future work and astronomical testing

Future work will be devoted to increase the bandwidth with a goal of 1-1.5GHz per sideband as well as to integrate the back end into the 1.2m Southern Millimeter Wave Telescope, a 115GHz observatory run by our group at the Cerro Calán National Observatory [DAS web page, 2013]. The upgrade will allow optimal observation of multiple CO lines. An example science case is the simultaneous observation of $^{12}$CO and $^{13}$CO in the galactic center. In this interstellar environment the intensity ratio $^{12}$CO/$^{13}$CO can be as low as 5 and the velocity range in excess of +/- 300 Km/s producing line widths of more than 250MHz. Having each line in one sideband, a high sideband rejection will be critical to be able to extract the exact line profiles all the way down to the continuum noise baseline.

## 6  Conclusion

A real-time, calibrated, digital sideband separating spectrometer has been built and tested. Two 500MHz IF channels were processed achieving a sideband rejection ratio better than 40 dB over the entire bandwidth and better than 50dB for most of the band. The bandwidth and spectral resolution is competitive for astronomical applications and the sideband rejection is 20 to 30dB better than current astronomical sideband separating receivers. This work demonstrates that the use of fast ADCs and FPGA based platforms to perform signal processing currently implemented by analog means can substantially increase the performance of sideband separating receivers.

### Acknowledgment
This work was financed by the Center of Excellence in Astrophysics and Associated Technologies (CATA) PBF06, the GEMINI-CONICYT Fund allocated to the project N° 32100003 and , and

Fondecyt Project No. 1121051. We thank Xilinx Inc. for the donation of FPGA chips and software licenses as well as the support of the CASPER collaboration. Special thanks to M. Morgan and J. Fisher for helping improving the manuscript.